# Comparing Popularity of Testing Careers among Canadian, Chinese, Indian Students

Luiz Fernando Capretz
Dept. of Electrical & Computer Eng.
Western University
London, Canada
lcapretz@uwo.ca

Pradeep Waychal
CRICPE
Western Michigan University
Kalamazoo, USA
pradeep.waychal@wmich.edu

Jingdong Jia
School of Software
Beihang University
Beijing, China
jiajingdong@buaa.edu.cn

*Abstract*—Despite its importance, software testing is, arguably, the least understood part of the software life cycle and still the toughest to perform correctly. Many researchers and practitioners have been working to address the situation. However, most of the studies focus on the process and technology dimensions and only a few on the human dimension of testing, in spite of the reported relevance of human aspects of software testing. Testers need to understand various stakeholders' explicit and implicit requirements, be aware of how developers work individually and in teams, and develop skills to report test results wisely to stakeholders. These multifaceted qualifications lend vitality to the human dimension in software testing. Exploring this human dimension carefully may help understand testing in a better way.

## I. INTRODUCTION

This study attempts to solve the basic problem of the human dimension in software testing, i.e., the lack of competent testing professionals, by trying to understand the unwillingness of computer and software engineering students across different geographies, and their reasons for not taking up testing careers. Waychal et al. [1], Deakin et al. [2], and Santos et al. [3] have studied the problem in Canada, Norway, and Brazil, respectively. The research question, therefore, is: why computer and software engineering students (henceforth referred to as students) across different parts of the world are reluctant to consider a career as software testers?

## II. METHODOLOGY

The study analyzed the opinions of 254 computer and software engineering undergrad students from three different countries (85 from Canada, 99 from China, and 70 from India) about their willingness to take up testing careers and the factors impacting their decisions.

The instrument selection was designed to understand the willingness of students to take up software testing careers and the reasons thereof. Specifically, the students were asked for the probability of their choosing testing careers by selecting one of the following choices: 'Certainly Yes,' 'Yes,' 'Maybe,' 'No,' and 'Certainly Not.' The study asked the respondents to provide open-ended but prioritized list of advantages and drawbacks, and open-ended rationale regarding their decisions on taking up testing careers. Since there has been limited prior research in the area, especially in the geographies that we were studying, we decided to use such a qualitative approach to investigate and understand the phenomena within their real-life context.

The objectives of the study were explained to students, and their responses to the survey were sought. They were assured that their responses would not influence course grades in any way and were offered an option of not disclosing their identities, gender and GPA. Table I presents the probabilities of students taking up testing careers, and the following subsections list advantages and drawbacks indicated by all respondents.

TABLE I. PROBABILITIES OF STUDENTS TAKING UP TESTING CAREERS

| Responses (254) | Canada (85) | China (99) | India (70) |
|---|---|---|---|
| Certainly Not | 31% | 24% | 14% |
| No | 27% | 0% | 31% |
| May be | 33% | 74% | 47% |
| Yes | 7% | 2% | 7% |
| Certainly Yes | 2% | 0% | 0% |

*Canada*: Only 9% of students were willing to be testers and 2% of them responded with 'Certainly Yes'. But a huge 58% of the students were not ready to take up testing careers, and 31% of them responded with 'Certainly No'. A significant 33% chose the 'May Be' option.

*China:* Most of the students (74%) chose the 'May Be' option. 24% were very clear that they would not take up the testing career and chose 'Certainly Not', and only 2% of students chose the 'Yes' option.

*India:* No student chose the 'Certainly Yes' option and only 7% of students chose the 'Yes' option. 14% of students selected 'Certainly Not' and 31% opted for the 'No' option. 47% percent of students were unsure and chose 'May Be'.

### A. Advantages of testing careers:

The responses from each country are analyzed and presented in Table II. Since we excluded advantages that were less than 5% (too small to consider) of the total advantages, the total in each column may not be 100%.

*Canada*: 21% of advantages recognized the testing jobs to be important and the same percentage found them to be easy, and 20% realized that there are more testing jobs. While 11% of advantages referred to the learning opportunities the testing careers offered, 9% asserted to testing having proper monetary rewards. Testing jobs were also seen as 'thinking' jobs (8%) and with the prospect of 'fun to break things's (7%).

*China*: 44% and 22% of advantages referred to testing being easy and offering more jobs, respectively. 13% of advantages indicated to proper monetary rewards for testing professionals. 'Learning opportunities' (8%), and 'fun to break things' (6%) were the next set of advantages.

*India:* Indian students' most voted advantage was testing being a 'thinking job' (38%). Its learning opportunities and importance fetched 30% and 14% of advantages, respectively. Easiness of the job polled just 9% of advantages.

TABLE II. PERCENTAGES OF SALIENT ADVANTAGES

| Advantages | Canada | China | India |
|---|---|---|---|
| Learning Opportunities | 11% | 8% | 30% |
| Important Job | 21% | | 14% |
| Easy Job | 21% | 44% | 9% |
| Thinking job | 8% | | 38% |
| More jobs | 20% | 22% | |
| Monetary benefits | 9% | 13% | |
| Fun to break things | 7% | 6% | |

### B. Drawbacks of testing career

The responses of students from each country are analyzed and presented in Table III. Since we excluded drawbacks that were less than 5% of the total drawbacks, the total in each column may not be 100%.

TABLE III. PERCENTAGES OF SALIENT DRAWBACKS

| Drawbacks | Canada | China | India |
|---|---|---|---|
| Second-class citizen | 17% | | 25% |
| Career development | | 6% | |
| Complexity | 6% | 37% | 24% |
| Tedious | 50% | 35% | 24% |
| Missing development | 15% | 7% | 15% |
| Less monetary benefits | | 6% | |
| Finding others' mistakes | 6% | | |
| No interest | | | 5% |

*Canada*: The most voted drawbacks for Canadian students were tediousness (50%), 'treatment as second-class citizens' (17%) and 'missing development' (15%). Complexity and finding mistakes of others polled 6% each.

*China*: The Chinese students' highest votes went to complexity (37%) and tediousness (35%) of testing jobs. 'Missing development' fetched 7%, and 'limited career development' and 'less monetary benefits' polled around 6% each.

*India:* The Indian students' most important drawbacks were treatment as 'second-class citizens' (25%) and testing being complex (24%) and tedious (24%). 15% referred to 'missing development' and 5% to a lack of personal interest.

The comparative study found that the positive aspects of pursuing testing careers (advantages) were very limited and included gaining experience (learning opportunities) and to the importance of job. The students didn't report the positive aspects of testing such as easy jobs, thinking jobs, better job prospects, and better monetary benefits, which this study discovered.

The negative aspects (drawbacks) considered by the comparative study included boredom (tediousness), 'no opportunities for writing code' ('miss development'), not creative (tediousness), and poor status and unrewarding ('second-class citizens').

### III. CONCLUSIONS AND IMPLICATIONS

This study analyzed the opinions of 254 computer and software engineering undergrad students in Canada, China, and India whether they would choose testing careers and what they felt were the advantages and drawbacks of the testing career.

The general empirical findings on the advantages as perceived by students from these countries do not seem to converge. While learning opportunities and easiness of the job are the common advantages, their percentages vary widely from 8% to 44%. While Indian students' major advantages are 'testing is a thinking job' and 'testing offers learning opportunities', for Chinese students it is the 'easiness and the large number of testing jobs', and for Canadians: 'easiness, importance and the large number of testing jobs'. In case of drawbacks, there is relatively a better convergence. The common drawbacks are tediousness, complexity, and 'missing development', although the range varies widely from 6% to 50%. 'Treatment as second-class citizen' figures high in the drawbacks of Canadian and Indian students.

The study has many implications for colleges, especially for computer and software engineering departments. Since testing courses can improve the perception of testing careers, colleges can introduce them in their curricula. They can regularly review the curricula by consulting their alumni and to ongoing research. Since testing offers additional jobs, the course can help colleges improve placement prospects of their students.

The testing curriculum needs to reflect the understanding that testers need to provide correct information to various stakeholders, and appreciate that testing is 'applied epistemology' grounded in 'cognitive psychology'. The faculty must dispel beliefs such as, testing is just mechanically running tests and comparing outputs with expected results. Instead, they should explain the importance of testing and the philosophy behind it and impress upon the students that any testing assignment and design of test cases can be very creative. They should develop testers who can understand different domains and the needs of users in those domains, to understand the developer mindset and anticipate mistakes that developers may be making as an individual and as a team, to test creatively and efficiently under the given constraints, and report the findings wisely to all stakeholders.

Additionally, the advantages and drawbacks of testing careers, as perceived by students, can help test managers and team leaders scale the challenge of recruiting test professionals. Understanding the common as well as country-specific advantages and drawbacks may help managers dealing with global teams. As emphasized before, software testing is a human activity [4] and testers, who willingly take up testing careers [5], can influence the quality of the final product.